\documentclass[aps,pre,preprint,groupedaddress]{revtex4}
\usepackage{graphicx}
\usepackage{latexsym}
\usepackage{amsmath,amssymb}

\begin{document}
%\preprint{}
%%%\begin{CJK*}{UTF8}{}

\title{
%Folding of the triangular lattice
%without the open edges:
%A modification of the folding rule
Crumpling transition of the triangular lattice
without open edges: effect of a modified folding rule
}

\author{Yoshihiro Nishiyama  }%%%(西山由弘)}
%\thanks{}
\affiliation{Department of Physics, Faculty of Science,
Okayama University, Okayama 700-8530, Japan}

\date{\today}

\begin{abstract}
Folding of the triangular lattice
in a discrete three-dimensional space
is investigated by means of the transfer-matrix method.
This model was introduced by
Bowick and co-workers as a discretized version of 
the polymerized membrane in thermal equilibrium.
The folding rule (constraint) is incompatible with the
periodic-boundary condition, and 
the simulation has been made under 
the open-boundary condition.
In this paper, we propose a modified constraint,
which is compatible with the periodic-boundary condition;
technically, the restoration of translational invariance
leads to a substantial reduction of the 
transfer-matrix size.
Treating the cluster sizes $L \le 7$,
we analyze the singularities of the crumpling transitions
for a wide range of the bending rigidity $K$.
We observe a series of the crumpling transitions at $K=0.206(2)$,
$-0.32(1)$, and $-0.76(10)$.
At each transition point, we
estimate the latent heat as $Q=0.356(30)$,
$0.08(3)$, and $0.05(5)$,
respectively.
\end{abstract}
\pacs{
82.45.Mp % Thin layers, films, monolayers, membranes Membranes, bilayers,
         % and vesicles
05.50.+q % Lattice theory and statistics (Ising, Potts, etc.) (see also
% 64.60.Cn Order-disorder transformations and statistical mechanics
         %  of model systems and
%  07.05.Tp Computer modeling and simulation
5.10.-a % Computational methods in statistical physics and
        % nonlinear dynamics (see also
%02.70.-c in mathematical methods in physics)
46.70.Hg % Membranes, rods and strings
%  05.10.Cc Renormalization group methods
% 75.10.Hk Classical spin models)
}
%\keywords{}

\maketitle

%%%\end{CJK*}

\section{\label{section1} Introduction}

At sufficiently low temperatures,
the polymerized membrane becomes flattened macroscopically
\cite{Nelson87}; see Refs.
\cite{Nelson89,Nelson96,Bowick01} for a review.
(The constituent molecules of the polymerized membrane 
have
a fixed connectivity, and the in-plane strain is subjected to
finite shear moduli.
This character is contrastive to 
that of the fluid membrane \cite{Canham70,Helfrich73}, which does not
support a shear.)
The flat phase is characterized by the long-range
orientational order of the surface normals.
It is rather peculiar that
such a
continuous (rotational) symmetry 
is broken spontaneously for such a two-dimensional manifold.
To clarify this issue,
a good deal of theoretical analyses have been reported so far.
However, it is still unclear 
whether the transition is critical 
\cite{Kantor86,Kantor87,Baig89,Ambjorn89,Renken96,Harnish96,Baig94,Bowick96b,Wheather93,Wheater96,David88,Doussal92,Espriu96}
or belongs to a discontinuous one with
an appreciable latent heat
\cite{Paczuski88,Kownacki02,Koibuchi04}.
Actually, in numerical simulations,
it is not quite obvious to
rule out the possibility of a weak-first-order transition
\cite{Kantor87b,Kantor87c};
see also Ref. \cite{Kownachi09}.

Meanwhile, a discretized version of the polymerized membrane
was formulated by Bowick and coworkers
\cite{Bowick95,Cirillo96,Bowick97}.
To be specific, they considered 
a sheet of the triangular lattice
embedded in a discretized three-dimensional space
(face-centered-cubic lattice); see Fig. \ref{figure9} (a).
(Even more simplified folding model, the so-called planar folding,
was studied in Refs. \cite{Kantor90,DiFrancesco94b,Cirillo96b}.)
Owing to the discretization,
the folding model admits an Ising-spin representation,
for which
a variety of techniques,
such as the mean-field theory and the transfer-matrix method,
are applicable.
A peculiarity of this Ising magnet is that 
the spin variables are subjected to a local constraint
(folding rule),
 which is incompatible with the periodic-boundary
condition.
%(Actually, the cylindrical surface is robust against deformation.)
Because of this difficulty, 
the open-boundary condition has been implemented
so far \cite{Bowick95,Bowick97,Nishiyama04,Nishiyama05}.
With 
the full diagonalization method, 
the finite clusters with the sizes $L \le 6$
were considered \cite{Bowick95,Bowick97}.
By means of the density-matrix renormalization group
(DMRG) \cite{White92,Peschel99},
the clusters with $L \le 29$ and $26$
were treated in 
Refs. \cite{Nishiyama04} and \cite{Nishiyama05}, respectively.
(The results are overviewed afterward.)

In this paper, 
we modify the local constraint,
aiming to implement the periodic-boundary condition,
and restore the translation invariance.
Technically, the restoration of translation invariance
admits a substantial reduction of the transfer-matrix size.
Taking the advantage, we treat 
the sizes up to $L=7$,
and analyze the singularities of crumpling transitions in detail.

The cluster-variation method (CVM)
(based on the single-hexagon approximation)
revealed a rich character of the discrete folding
\cite{Cirillo96,Bowick97}.
According to CVM,
there appear the totally flat, 
octahedral, tetrahedral, and piled-up phases, as the bending rigidity
$K$ changes from $\infty$ to $-\infty$.
%%%%%%%%%%%%%%%%%%%%%%%%%%%%%%%%%%%%%%%%%%%%%%%%%%%%%%%%%%%%
Namely, in respective phases,
the triangular-lattice sheet crumples up to form a 
octahedron, a tetrahedron, and a triangular plaquette;
see Fig. 3 of Ref. \cite{Bowick95}.
%%%%%
(This picture is based on 
a single-hexagon approximation of CVM.
Beyond a mean-field level, 
the thermal undulations may be 
induced, particularly,
in the vicinity of the transition point,
disturbing the shape of the crumpled sheet significantly.)
%%%%%%%%%%%%%%
More specifically,
the crumpling transitions separating these phases are
estimated as
$K=0.185$,    $-0.294$, and    $-0.852$;
hereafter, we abbreviate the set of parameters
as $(0.185,-0.294,-0.852)$.
At each transition point, the latent heat is estimated as
$(0.229 ,  0.14 , 0)$;
namely, the third transition is continuous
according to CVM.
On the one hand,
the
DMRG simulation \cite{Nishiyama04,Nishiyama05}
indicates
the transition point 
$[ 0.195(2) , -0.32(1) , -0.76(1) ]$ with the latent heat
$[ 0.365(5) ,  0.04(2) , 0.03(2) ]$;
the character of the third transition point
is still controversial.

The rest of this paper is organized as follows.
In Sec. \ref{section2}, 
we propose
a modified folding rule (\ref{modified});
the transfer-matrix formalism
\cite{Bowick95} is explicated in the Appendix.
In Sec. \ref{section3}, we present the numerical results.
The singularities
of the crumpling transitions are analyzed in detail.
In Sec. \ref{section4},
we present the summary and discussions.

\section{\label{section2}
A modification of the folding rule}

In this section, we present a modified folding rule, 
Eq. (\ref{modified}).
As mentioned in the Introduction, 
the folding rule, enforced by the
prefactors in Eq. (\ref{TM}), is too restrictive
to adopt the periodic-boundary condition.
So far,
the numerical simulation has been performed under the
open-boundary condition
\cite{Bowick95,Bowick97,Nishiyama04,Nishiyama05}.

To begin with, we outline the transfer-matrix formalism;
an explicit algorithm is presented in 
the Appendix.
According to Ref. \cite{Bowick95}, through
a dual transformation,
the triangular-lattice folding reduces to
an Ising model on the hexagonal lattice, Fig. \ref{figure9} (a).
A drawing of a transfer-matrix strip is presented in Fig. \ref{figure9} (b).
A peculiarity of this reduced Ising model is that
the spins surrounding each hexagon are subjected to a constraint.
(The constraint originates from the folding rule.)
To be specific, 
the
prefactors
$U_j V_j(=0,1)$
of the transfer-matrix element (\ref{TM})
restrict the configuration space.
As mentioned in the Introduction,
this constraint is incompatible with the periodic-boundary condition.
(For instance, a cylindrical paper supports a large strain,
whereas an open paper is flexible.)

%Actually, cylindrical paper supports a large strain,
%whereas open paper is freely folded.

Aiming to restore the translational invariance,
we modify the prefactors.
We replace Eq. (\ref{TM}) with
\begin{equation}
\label{modified}
\frac{1}{L}
\sum_{l=1}^L (\prod_{j \ne l} U_j V_j) 
     [(1-p)U_l V_l + p]
\exp [ -\sum_{k \ne l} H_k (K) - H_l(K')]
.
\end{equation}
Here, the parameter $L$ denotes the system size,
and the explicit formulas for the constraint $U_j V_j$
and the elastic energy $H_k(K)$ are shown in the
Appendix.
%%%%%%%%%%%%%%%%%%%%%%%%%%%%%%%%%%%%%%
As compared with the original form
(\ref{TM}), our modified expression
(\ref{modified}) has a defect at $l$,
where the folding-rule constraint is released.
Because the defect is distributing uniformly
by the summation $\sum_{l=1}^L$
and the normalization factor $1/L$,
the translational invariance is maintained.
%In Eq. (\ref{modified}),
%the prefactor at the position $l$
%has a modified expression $(1-p)U_l V_l + p$.
%That is,
%we inject a defect,
%at which
%the local constraint is released by modifying
%the prefactor $U_lV_l$.
The parameter $K'$ describes the local elastic constant at the defect.
The probability of the defect is controlled by the parameter $p$;
at $p=0$, the original constraint $U_l V_l $ recovers, whereas at $p=1$, 
the constraint disappears $U_l V_l \to 1$.
We stress that a single defect does not alter the thermodynamic
(bulk) properties.
As a byproduct,
two tunable parameters $p$ and $K'$
are available.
The parameters are adjusted to
\begin{equation}
\label{defect_parameters}
(p,K')=(0.7, 1.5 K)
  .
\end{equation}
A justification of this choice is
given in
Sec. \ref{section3_2}
(Fig. \ref{figure2}).

\section{\label{section3}Numerical results}

In this section, we present the simulation results.
We employed the transfer-matrix method (Appendix)
with a modified folding rule, Eq. (\ref{modified}).
The numerical diagonalization was performed 
within a
subspace specified by the wave number $k=0$ and the parity even.
(In a preliminary survey,
we confirmed that the dominant-eigenvalue (thermal equilibrium) state
belongs to this subspace.)
Here, we make use of the spin-inversion symmetry 
$\sigma_i,z_i \to \pm \sigma_i , \pm z_i$.
(This symmetry group originates from the overall rotation of the crumpled triangular-lattice sheet.)
We stress that 
the wave number $k$ makes sense owing to the
restoration of
the translational invariance 
(\ref{modified}).
%As mentioned in the above section,
%the defect parameters are adjusted to Eq. (\ref{defect_parameters}).
%A justification of this choice is given
%in Sec. \ref{section3_2}.

\subsection{\label{section3_1}
Crumpling transitions: A preliminary survey}

\begin{figure}
\includegraphics[width=100mm]{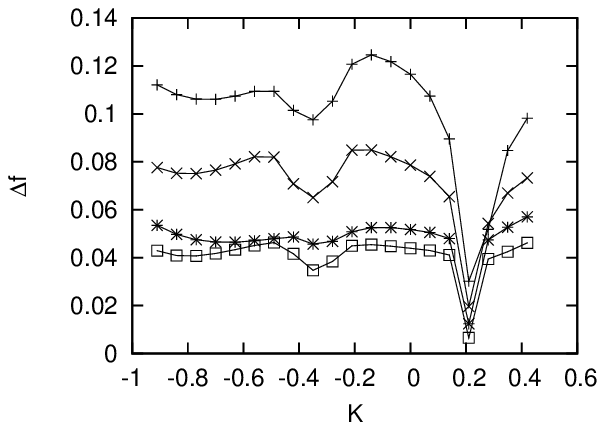}%
\caption{\label{figure1}
The free-energy gap (\ref{energy_gap}) is plotted for the bending rigidity $K$
and the system size $L=$($+$) 4, ($\times$) 5, ($*$) 6, and ($\Box$) 7.
The data indicate singularities (crumpling transitions)
around $K \approx 0.2$, $-0.3$, and $-0.8$;
detailed analyses of each transition are made
in Figs. \ref{figure3}-\ref{figure8}.
}
\end{figure}

In Fig. \ref{figure1}, 
we plot the free-energy gap
\begin{equation}
\label{energy_gap}
\Delta f= f_2-f_1
   ,
\end{equation}
for the bending rigidity $K$.
Here, the free energy per unit cell
is given by
$f_i=- \ln \Lambda_i /(2L)$ with
the (sub)dominant eigenvalue  $\Lambda_{1(2)}$
of the transfer matrix.
[Here, the unit cell stands for a triangle of
the original lattice 
rather than a hexagon of the dual lattice; see Fig. \ref{figure9} (a).]

As mentioned in the Introduction,
the triangular-lattice sheet becomes crumpled, as the rigidity $K$ 
changes from $\infty$ to $-\infty$.
In Fig. \ref{figure1},
we see a number of signatures
of the crumpling transitions
around $K \approx 0.2$, $-0.3$, and $-0.8$.
(Note that the closure of the excitation gap
indicates an onset of phase transition.)
%%%%%%%%%%%%%%%%%%%
On the one hand,
the CVM analysis \cite{Cirillo96,Bowick97}
predicts a series of crumpling transitions at
$(0.185,-0.294,-0.852)$.
The results appear to be consistent with
those of Fig. \ref{figure1},
suggesting that
the excitation-gap closure
indicates a location of the crumpling transition.
%%%albeit these signatures are subtle. 
%%%%%%%%%%%%%%%
%%%%%%%%%%%%%%%
%This behavior is consistent with the preceding studies 
%(see the Introduction).
Detailed analyses of each singularity are made in
Sec. \ref{section3_3} and Sec. \ref{section3_4}.

\subsection{\label{section3_2}
Simulation at $p=1$ and $K'=0$}

As a comparison, we provide a simulation result,
setting
the defect parameters to 
$p=1$ and $K'=0$ tentatively.
This parameter set 
has an interpretation that
a rupture (pair of open edges)
distributes 
uniformly along the
transfer-matrix strip.
(This situation is an extention of
the 
open-boundary condition,
for which the rupture is static.)

\begin{figure}
\includegraphics[width=100mm]{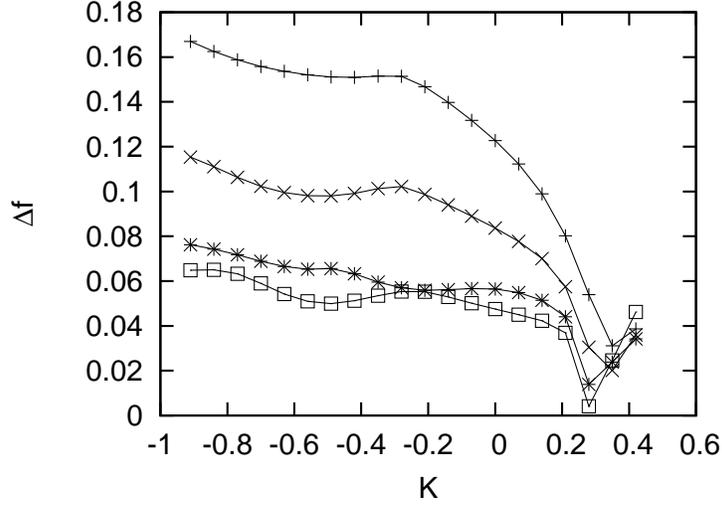}%
\caption{\label{figure2}
The free-energy gap (\ref{energy_gap}) is plotted for the bending rigidity $K$
and the system size $L=$($+$) 4, ($\times$) 5, ($*$) 6, and ($\Box$) 7.
Tentatively, the defect parameters [Eq. (\ref{modified})] are
set to $p=1$ and $K'=0$.
%, corresponding to a case of distributing ruptures over the membrane.
The data
appear to be scattered, as compared to those of Fig. \ref{figure1}.
This result
indicates that the finite-size behavior
is improved by adjusting the defect parameters.
}
\end{figure}

In Fig. \ref{figure2}, we present the free-energy gap $\Delta f$
for the bending rigidity $K$;
the range of $K$ is the same as that of Fig. \ref{figure1}.
%Clearly, the data of Fig. \ref{figure2} 
%are scattered, as compared to those
%of Fig. \ref{figure1}.
The signatures of crumpling transitions
in Fig. \ref{figure2}
are less clear, as compared to those 
of Fig. \ref{figure1}.
%As a matter of fact,
%signatures of the crumpling transitions
%become unclear.
This result
indicates that 
the choice of the defect parameters
affects the finite-size behavior.
In the preliminary stage, we surveyed
a parameter space of $p$ and $K'$,
and arrived at a conclusion that the
above choice, Eq. (\ref{defect_parameters}),
is an optimal one.
Note that these parameters are the byproduct of 
the modification of the folding rule,
Eq. (\ref{modified}).
Here, we make use of these redundant parameters
so as to improve the finite-size behavior,
aiming to
take the thermodynamic limit reliably.

\subsection{\label{section3_3}
Crumpling transition in $K>0$: 
Analysis of the
latent heat via the Hamer method}

In Sec. \ref{section3_1},
we observed a series of crumpling transitions.
In this section, we analyze the singularity of a transition
in the $K>0$ side.

To begin with, we determine the transition point.
In Fig. \ref{figure3},
an approximate transition point $K_c(L)$ is plotted
for $1/L^2$.
Here, the approximate transition point
minimizes the free-energy gap
\begin{equation}
\label{transition_point}
\partial_K \Delta f |_{K=K_c(L)}=0
.
\end{equation}
The least-squares fit yields an estimate $K_c=0.20617(70)$.
Similarly, as for $5 \le L \le 7$, we obtain
an estimate $K_c=0.20506(99)$.
A discrepancy between these results may be an indicator 
of possible systematic error.
The systematic error 
appears to be comparable to the fitting error.
Regarding both errors as the sources of error margin,
we obtain
\begin{equation}
\label{transition_point_1}
K_c=0.206(2)
.
\end{equation}

Based on the transition point $K_c(L)$,
we estimate the amount of latent heat.
According to Ref. \cite{Hamer83},
the low-lying eigenvectors of the transfer matrix
contain information 
on
the latent heat.
We explain the underlying idea,
and present the scheme explicitly.
At the discontinuous (first-order) transition point,
the low-lying spectrum of the transfer matrix
exhibits a level crossing, and the 
discontinuity (sudden drop)
of the slope 
reflects a release of the latent heat.
However, the finite-size artifact (level repulsion)
smears out the singularity.
According to Ref. \cite{Hamer83},
regarding the low-lying levels as nearly degenerate,
one can resort to the 
perturbation theory of the degenerated case,
and 
calculate the level splitting (discontinuity of slope)
explicitly.
To be specific,
we consider the matrix
\begin{equation}
\label{perturbation_matrix}
V=
\left(   
\begin{array}{cc}
V_{1 1}  &  V_{1 2}  \\
V_{2 1}  &  V_{2 2}
\end{array}
\right)
       ,
\end{equation}
with $V_{ij}=\langle i |  \partial_K T | j \rangle$.
The matrix $T$ denotes the transfer matrix; 
namely, the matrix element $\partial_K T$
is given by a product of 
Eq.
(\ref{TM}) and
Eq. 
(\ref{Hamiltonian_matrix})
 with $K$ dropped.
The bases, $|1\rangle $ and $| 2 \rangle$,
are 
the (nearly degenerate) eigenvectors of $T$
with the eigenvalues
$\Lambda_{1,2}$, respectively.
The states 
$  \{ | i \rangle \} $ are normalized so as to satisfy 
$\langle i  | T |  i \rangle  =1$.
According to the perturbation theory,
the eigenvalues of Eq. (\ref{perturbation_matrix}) 
yield the level-splitting slopes due to $K$.
Hence, the latent heat 
(per unit cell)
is given by
a product of
this discontinuity and the elastic constant
\begin{equation}
\label{latent_heat}
Q (L) =       | K_c (L) |
\sqrt{  (V_{11}-V_{22})^2+4V_{12}V_{21} }   
          \frac{1}{2L}
,
\end{equation}
for the system size $L$.

In Fig. \ref{figure4},
we plot the latent heat (\ref{latent_heat}) for $1/L^2$.
The least-squares (linear) fit yields an estimate $Q=0.356(12)$ in the thermodynamic limit.
Similarly, we obtain $Q=0.3774(93)$ for $5 \le L \le 7$.
Again, the systematic error appears to be comparable to the fitting error.
We estimate 
\begin{equation}
\label{latent_heat_1}
Q=0.356(30)
.
\end{equation}

This is a good position to address a few remarks.  %%%%%  a remark.
First,
the latent heat, Eq. (\ref{latent_heat_1}),
agrees with that of DMRG \cite{Nishiyama04} (see the Introduction),
whereas the transition point, Eq. (\ref{transition_point_1}), lies out of the
error margin.
This discrepancy may indicate an existence of systematic
error as to the determination of $K_c$.
A peculiarity \cite{DiFrancesco94b}
of this transition is that in the $K>K_c$ side,
the system becomes completely flattened;
namely, there exist no thermal undulations, as if
the system is in the low-temperature limit $K\to\infty$.
This peculiarity may gives rise to a bias to $K_c$.
(As a matter of fact, in Ref. \cite{Nishiyama04},
a pronounced hysteresis was observed.)
On the contrary, 
we confirmed that 
the ambiguity of $K_c$ does not influence the latent heat $Q$ very much.
(Because of this $K_c$ independence, the result $Q$ is reliable,
as the above-mentioned consistency with DMRG suggests.)
In the next section, 
the remaining two transitions
are analyzed in a unified manner.
%%%%%%%%%%%%%%%%%%%%%%%%%%%%%%%%%%%%%%%
Second,
we consider the $1/L^2$ extrapolation scheme.
The finite-size data converge rapidly
to the thermodynamic limit
around the first-order transition point,
because the correlation length (typical length scale)
$\xi$ remains finite.
Hence, the dominant system-size corrections
should be described by
$1/L^2$ (rather than $1/L$).

\subsection{\label{section3_4}
Crumpling transitions in $K<0$}

In this section, we analyze the remaining transitions
in the $K<0$ side.

First, we analyze the transition around $K \approx -0.3$.
In Fig. \ref{figure5}, we plot the transition point (\ref{transition_point})
for $ 1/L^2 $.
The least-squares fit to these data yields an estimate
$K_c=-0.320(12)$.
Similarly, we obtain $K_c=-0.316(28)$ for $5 \le L \le 7$.
The systematic error appears to be negligible,
as compared to the fitting error.
Considering the latter as the source of error margin, we estimate the transition point as
\begin{equation}
K_c=-0.32(1)
.
\end{equation}
In Fig. \ref{figure6}, we plot the latent heat (\ref{latent_heat})
for $1/L^2$.
The least-squares fit yields $Q=0.077(15)$. 
Similarly, for $5 \le L \le 7$, we obtain $Q=0.058(28)$.
The fitting and systematic errors are comparable.
Considering them as the sources of error margin,
we estimate the latent heat as
\begin{equation}
Q=0.08(3)
.
\end{equation}

Second, we turn to the analysis of the transition around $K \approx -0.8$.
In Fig. \ref{figure7}, we plot the transition point (\ref{transition_point})
for $ 1/L^2 $.
The least-squares fit to these data yields an estimate
$K_c= -0.76(10)$.
Similarly, we obtain $K_c= -0.72(24)$ for $5 \le L \le 7$.
The fitting error dominates the systematic error.
Neglecting the latter,
we obtain
\begin{equation}
K_c = -0.76(10)
.
\end{equation}
In Fig. \ref{figure8}, we plot the latent heat (\ref{latent_heat})
for $1/L^2$.
The least-squares fit yields $Q= 0.049(51)$. 
Similarly, for $5 \le L \le 7$, we obtain $Q=0.01(11)$.
Again, the systematic error appears to be negligible.
We estimate
the latent heat as
\begin{equation}
\label{latent_heat_3}
Q = 0.05(5)
.
\end{equation}

%The result (\ref{latent_heat_3}) does not exclude
%a possibility of a continuous (critical) transition with $Q=0$.
%In order to examine a possibility
%of $Q=0$,
%we calculated the reciprocal correlation length
%$1/\xi = \ln (\Lambda_1/\Lambda_2) |_{K=K_c(L)}  $.
%According to the finite-size-scaling theory,
%the reciprocal correlation length
%vanishes in the form $\propto 1/L$ 
%at the critical point.
%The least-squares fit to the data,
%$1/L$-$1/\xi$, yields an estimate $1/\xi=0.13(13) $
%in the thermodynamic limit.
%This supplementary result may suggest a tendency toward a discontinuous 
%transition, although the amount of error margin
%still prevents a definite answer.

\begin{figure}
\includegraphics[width=100mm]{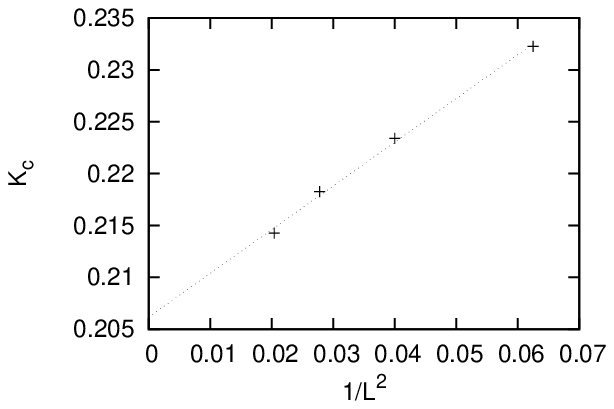}%
\caption{\label{figure3}
The crumpling transition $K \approx 0.2 $ observed in 
Fig. \ref{figure1} is analyzed in detail.
The transition point $K_c(L)$ (\ref{transition_point}) is plotted for $1/L^2$.
The least-squares fit yields an estimate $K_c=0.20617(70)$.
}
\end{figure}

\begin{figure}
\includegraphics[width=100mm]{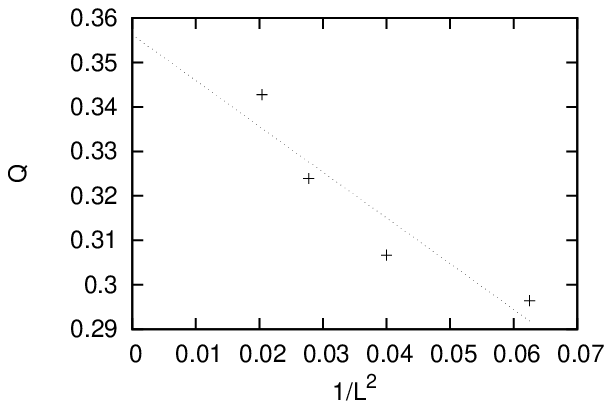}%
\caption{\label{figure4}
The latent heat $Q(L)$
(\ref{latent_heat}) is plotted for $1/L^2$.
The least-squares fit yields an estimate $Q=0.356(12)$.
}
\end{figure}

\begin{figure}
\includegraphics[width=100mm]{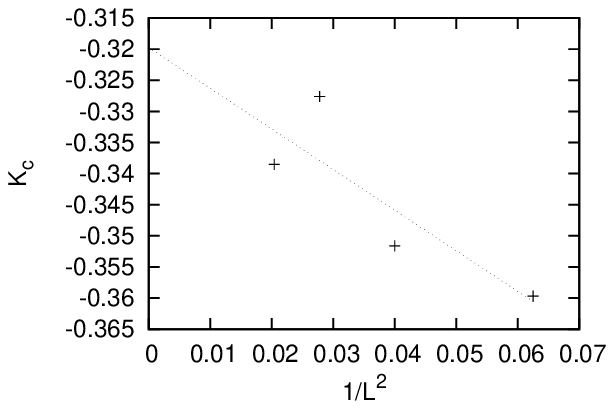}%
\caption{\label{figure5}
The crumpling transition $K \approx -0.3 $ observed in 
Fig. \ref{figure1} is analyzed in detail.
The transition point $K_c(L)$ (\ref{transition_point}) is plotted for $1/L^2$.
The least-squares fit yields an estimate $K_c=-0.320(12)$.
}
\end{figure}

\begin{figure}
\includegraphics[width=100mm]{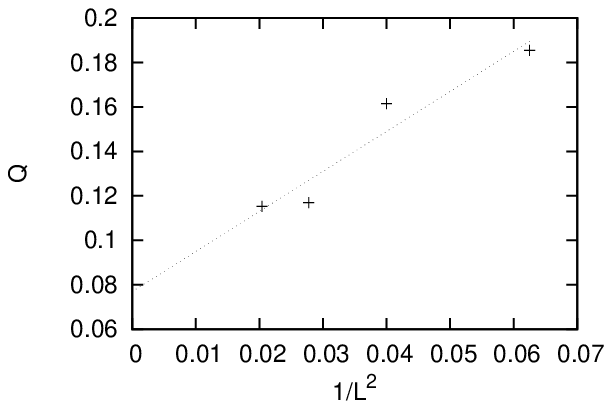}%
\caption{\label{figure6}
The latent heat $Q(L)$
(\ref{latent_heat}) is plotted for $1/L^2$.
The least-squares fit yields an estimate $Q=0.077(15)$.
}
\end{figure}

\begin{figure}
\includegraphics[width=100mm]{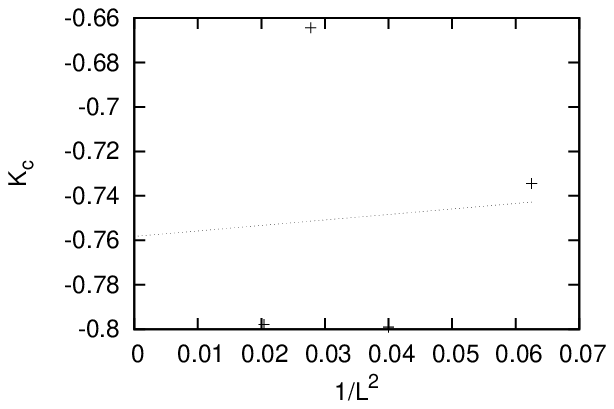}%
\caption{\label{figure7}
The crumpling transition $K \approx -0.8 $ observed in 
Fig. \ref{figure1} is analyzed in detail.
The transition point $K_c(L)$ (\ref{transition_point}) is plotted for $1/L^2$.
The least-squares fit yields an estimate $K_c = -0.76(10)$.
}
\end{figure}

\begin{figure}
\includegraphics[width=100mm]{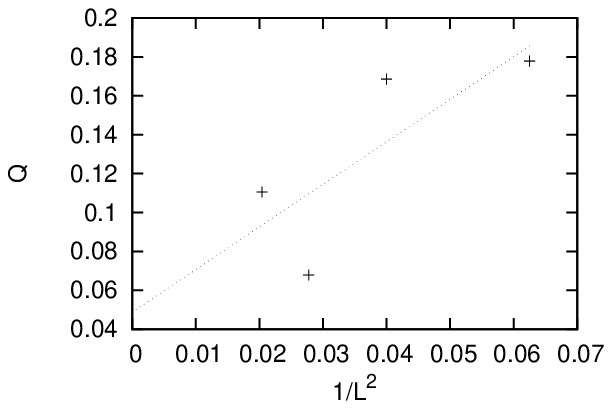}%
\caption{\label{figure8}
The latent heat $Q(L)$
(\ref{latent_heat}) is plotted for $1/L^2$.
The least-squares fit yields an estimate $Q= 0.049(51)$.
}
\end{figure}

Last, we consider a shaky character of
Figs. \ref{figure3}-\ref{figure8}
(in particular, Figs. \ref{figure5}-\ref{figure8}).
Such a shaky character
is an artifact due to
the cluster size.
In the preliminary stage,
we surveyed the planar folding
\cite{Kantor90,DiFrancesco94b,Cirillo96b} for
considerably large system sizes $L \le 14$;
the configuration space of the planar folding
is much restricted.
As a result, we found that 
the finite-size behavior is irregular
with respect to $L$;
this irregularity 
is an obstacle to making an extrapolation.
Roughly speaking,
the finite-size behavior is categorized by
$L=0$, $1$, and $2$ mod $3$.
Although the enlarged configuration space
suppresses this irregularity,
a slight irregularity for $L=6(=0$ mod $3)$
still remains.
A slight bump of Figs. \ref{figure5} and \ref{figure7}
may be due to this irregularity.
Hence, we consider that the deviations
(seemingly curved plots)
in Figs. \ref{figure3}-\ref{figure8}
are not systematic ones.
Rather, considering them as a source of errors,
we estimate the error margin 
by making two independent extrapolations
for different sets of system sizes.

%%This supplementary result supports
%%the above claim (\ref{latent_heat_3})
%%that the transition is discontinuous;
%%the error margin prevents a conclusive answer.

%A comment is in order,
%the present result $L$ at $L$ does not provide
%a conclusive evidence that the transition is discontinuous
%The singularity would not be captured L8
%\cite{Bowick97}

\section{\label{section4}
Summary and discussions}

We proposed 
a modified folding rule, Eq. (\ref{modified}),
which enables us to simulate the
discrete folding without the open edges.
By means of the transfer-matrix method,
we investigated a series of the crumpling transitions.
We estimate the transition point and 
the latent heat as
$[0.206(2),-0.32(1),-0.76(10)]$
and $[0.356(30),0.08(3),0.05(5)]$, respectively.
Our result agrees with the preceding CVM and DMRG results.
In particular, our result is in quantitative agreement with that of DMRG.

According to Ref. \cite{Bowick97},
the third singularity (around $K \approx -0.8$) is
so subtle that
it could not be captured until $L=8$.
On the contrary, our data in Figs. \ref{figure1} and \ref{figure7}
succeeds in detecting 
a signature of a crumpling transition
even for $ 4 \le L \le 7$.
As anticipated, the restoration of the translation invariance
leads to an improvement of
the finite-size behavior.

As mentioned in the Introduction,
it is still unclear 
whether the third
transition is continuous
\cite{Bowick97}
or
belongs to a weak-first-order transition
\cite{Nishiyama05}.
The present result (\ref{latent_heat_3})
does not exclude a possibility of a continuous transition.
According to Ref. \cite{Bowick97},
around $K \approx -0.8$, 
through a truncation of the
configuration space,
the discrete folding 
reduces to a simplified version of the folding model,
the so-called
planar
folding
\cite{Kantor90,DiFrancesco94b,Cirillo96b},
which exhibits a continuous transition in the $K<0$ side.
An examination of this truncation process
may provide valuable information on 
the nature of this phase transition.
This problem will be addressed in the future study.

%In order to settle this longstanding issue,
%further consideration would be desirable.

%In Eq. (),
%we introduced a (propagating) defect, at which the folding-rule constraint 
%is released.
%It would be tempting to ask how the injection of defects
%influences the phase diagram.
%In Ref. , it is argued that by the randomness,
%the crumpling transition turns into a continuous one,
%and a rich phase diagram emerges.
%Similarly, the injection of defects may alter the singularity
%of the crumpling transition significantly.
%This problem will be addressed in the future study.

%\begin{acknowledgments}
%This work was supported by a Grant-in-Aid 
%from Monbu-Kagakusho, Japan
%(No. 18740234).
%\end{acknowledgments}

\appendix*

\section{\label{appendix}
Transfer-matrix formalism for the discrete folding}

\begin{figure}
\includegraphics[width=100mm]{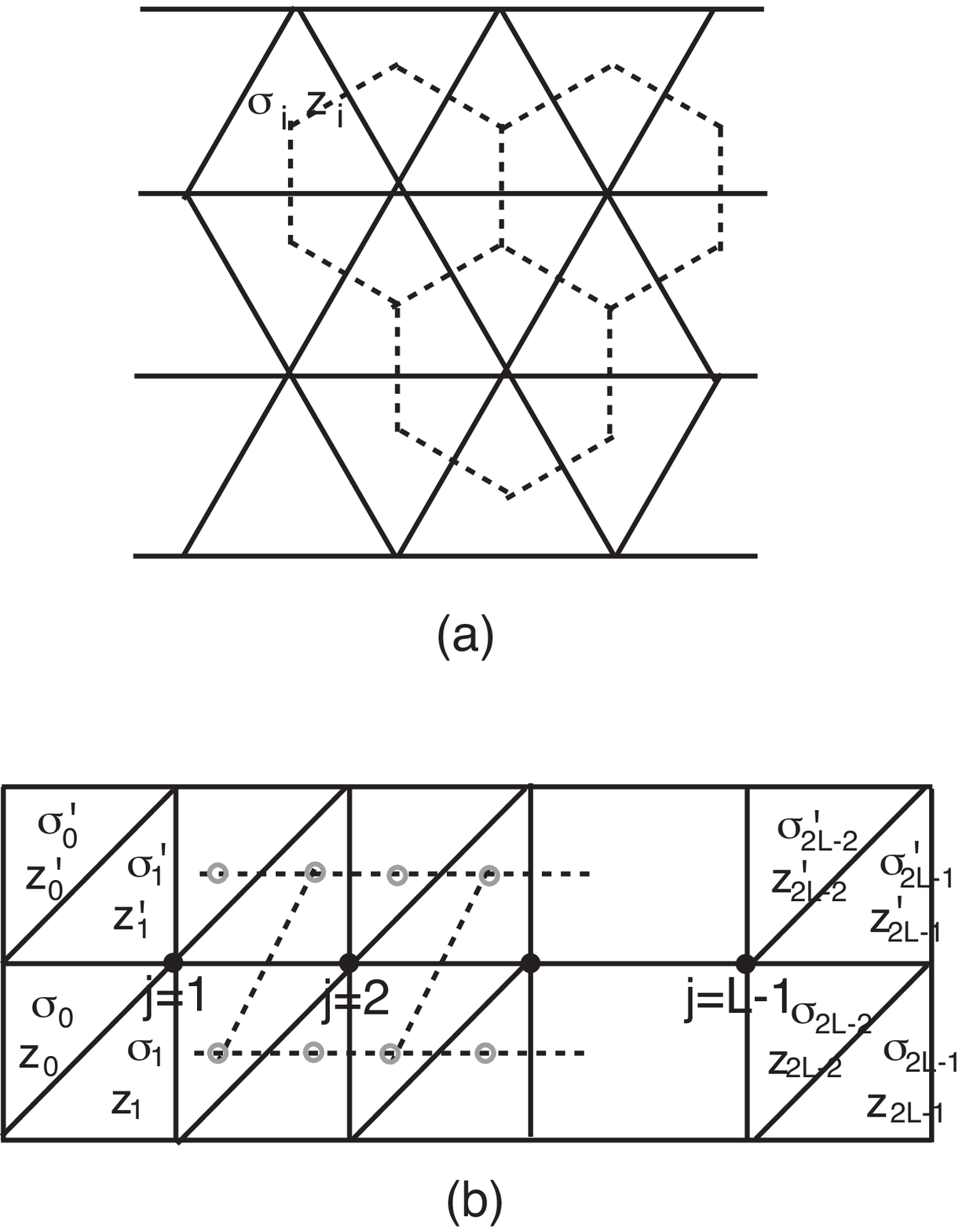}%
\caption{\label{figure9}
(a)
We consider a discrete folding of the triangular lattice.
In order to specify the fold angle,
we place two types of Ising variables $\{ z_i, \sigma_i \}$
on each triangle $i$ rather than at each joint (gauge rule \cite{Bowick95}).
Hence, hereafter, we consider a spin model
on the dual (hexagonal) lattice.
(b)
Construction of the transfer matrix.
The row-to-row statistical weight yields the transfer-matrix element,
Eq. (\ref{TM}).
So far, the open-boundary condition has been imposed.
Here, we restore the translational invariance
by using a modified folding rule (\ref{modified}).
}
\end{figure}

In this Appendix,
we present the transfer-matrix formalism for the discrete folding
\cite{Bowick95}.
%a modification of the folding rule 
%(\ref{}) is proposed in Sec. .
Before commencing a mathematical description,
we explicate a basic feature of the discrete folding.
We consider a sheet of
the triangular lattice, Fig. \ref{figure9} (a).
Along the edges, the sheet folds
up.
The fold angle $\theta$ along the edges is discretized 
into four possibilities, namely,
``no fold" ($\theta=\pi$),
``complete fold" ($\theta=0$),
``acute fold" [$\theta={\rm arccos}(1/3)$], and
``obtuse fold" [$\theta={\rm arccos}(-1/3)$];
in other words, 
the triangular lattice is embedded in the face-centered-cubic lattice
\cite{Bowick95}.

The above discretization leads to an Ising-spin representation.
The mapping,
the so-called gauge rule, reads as follows \cite{Bowick95}.
We place two types of Ising variables $\{ \sigma_i , z_i \}$
at each triangle $i$
(rather than each joint); see Fig. \ref{figure9} (a).
Hence, hereafter, we consider the a spin model on the
dual (hexagonal) lattice.
The gauge rule sets the joint angle between the adjacent triangles.
That is,
provided that the $z$ spins are antiparallel ($z_1 z_2=-1$) for a pair of adjacent neighbors,
the joint angle is either an acute or obtuse fold.
Similarly, if $\sigma_1 \sigma_2=-1$ folds, the relative angle is either a complete or obtuse fold.
The spins are subjected to a constraint (folding rule); 
The prefactors $U_j V_j$ of
the transfer-matrix element (\ref{TM}) 
enforces the constraint.

As a consequence, the discrete folding reduces to a two-component Ising model
on the hexagon lattice.
Hence, the transfer-matrix strip looks like that drawn in Fig. \ref{figure9} (b).
The row-to-row statistical weight $T_{ \{\sigma_i,z_i\},\{\sigma'_i,z'_i\}}$
yields the transfer-matrix element.
The transfer-matrix element for the strip width $L$ is 
given by \cite{Bowick95}
\begin{equation}
\label{TM}
T_{\{z_i,\sigma_i\},\{z'_i,\sigma'_i\}}=
(\prod_{j=1}^{L} U_j V_j) \exp(-H/T)           % periodic !!
,
\end{equation}
with
\begin{equation}
U_j=\delta(\sigma_{2j-2}+\sigma_{2j-1}+\sigma_{2j}+\sigma'_{2j-1}+\sigma'_{2j}+\sigma'_{2j+1}
\ mod\ 3,0)
,
\end{equation}
and
\begin{equation}
V_j=\prod_{c=1}^2 \delta(\alpha_c(z_{2j},z_{2j-1},z_{2j-2},z'_{2j-1},z'_{2j},z'_{2j+1})\ mod\ 2,0)
.
\end{equation}
The factors $\{U_j,V_j\}$ enforce the constraint (folding rule) as to the spins
surrounding each hexagon.
Here, 
$\delta(m,n)$ denotes Kronecker's symbol, and
$\alpha_c$ is given by
\begin{equation}
\alpha_c(z_1,\dots,z_6)=
\sum_{i=1}^6 \frac{1}{2}(1-z_iz_{i+1})\delta(   %%c_0+
\sum_{j=1}^i \sigma_j  \ mod\ 3,0) .
\end{equation}
The Boltzmann factor $\exp ( -H/T ) $ is due to the bending-energy cost $H$.
Hereafter, we choose the temperature $T$ as a unit of energy; namely, we set $T=1$.
As usual, the bending energy is given by the inner product 
$\cos \theta_{ij}$
of the surface normals of adjacent triangles.
Hence, the bending energy is given by a compact formula
\begin{equation}
\label{Hamiltonian_matrix}
H=\sum_k H_k  (K) .
\end{equation}
Here, the index $k$ specifies each hexagon as in Eq. (\ref{TM}).
The local energy of each hexagon $H_k$ is given by
\begin{equation}
H_k  (K) =
-0.5 \sum_{i=1}^6 K \cos \theta_{i,i+1}
= 
-0.5 \sum_{i=1}^6 \frac{1}{3}K\sigma_i \sigma_{i+1}
               (1+2z_i z_{i+1})
,
\end{equation}
with the bending rigidity $K$.
Here, 
the summation $\sum_{i=1}^6$ runs over all vertices around the hexagon $k$.
(The overall factor $0.5$ compensates the duplicated sum.)

%\begin{equation}
%H=-0.5 \sum_{\langle ij\rangle} K \cos \theta_{ij}=
%-0.5\sum_{\langle ij\rangle} \frac{1}{3} K \sigma_i\sigma_j(1+2 z_iz_j)
%\end{equation}
%with the bending rigidity $K$. Here,
%the summation $\sigma_{\langle ij\rangle}$ suns over all possible nearest-neighbor parts
%around each hexagon.
%(The overall factor $0.5$ compensates the duplicated sum.)


\end{document}